# 3D Data Long-Term Preservation in Cultural Heritage

Nicola Amico (*PRISMA srl*)

Achille Felicetti (*PIN scrl*)

This report was prepared under the auspices of the eArchiving Initiative. eArchiving is funded by the European Union's Digital Europe Programme. It is operated by the E-ARK Consortium led by the Austrian Institute of Technology (AIT) under a service contract with the European Commission, contract number LC-01905904.

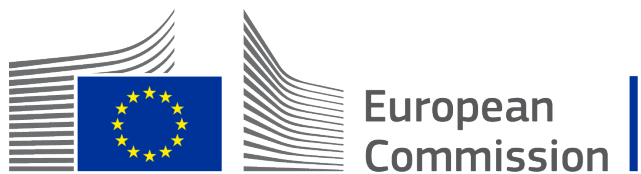

**Executive summary**

The report explores the challenges and strategies for preserving 3D digital data in cultural heritage. It discusses the issue of technological obsolescence, emphasising the need for sustainable storage solutions and ongoing data management strategies. Key topics include understanding technological obsolescence, the lifecycle of digital content, digital continuity, data management plans (DMP), FAIR principles, and the use of public repositories. The report also covers the importance of metadata in long-term digital preservation, including types of metadata and strategies for building valuable metadata. It examines the evolving standards and interoperability in 3D format preservation and the importance of managing metadata and paradata. The document provides a comprehensive overview of the challenges and solutions for preserving 3D cultural heritage data in the long term.

**Table of contents**

**Introduction**

This report addresses the complexities of technological obsolescence, exploring its implications for preserving cultural heritage and the strategies needed to navigate this evolving landscape.

In digital heritage, effective management and preservation of digital data are crucial. Issues such as file corruption, media obsolescence, and inadequate metadata must be addressed, alongside data migration when software becomes outdated and thorough data curation to aid current and future researchers in searching, citing, and reusing historical data.

Merely archiving or backing up project data is not enough for long-term preservation. It is essential to ensure that primary data remain reusable, compatible with evolving operating systems, and accompanied by comprehensive metadata detailing their creation and history [1].

Despite the advantage of heritage datasets being "born digital," they are still susceptible to loss if file associations and metadata are not properly maintained. The large volume of data generated from digital projects and the often limited understanding of file associations among project members jeopardise the future reuse of archaeological data if not well-organised or curated. Enhancing workflows to include both metadata authorship and preservation is vital to prevent information loss and digital data obsolescence.

Particularly, the long-term preservation of 3D datasets requires maintaining each file in a usable and uncorrupted state. Files undergo several modifications, changing formats during the creation of the final scan or 3D model, known as an asset. This preservation process necessitates archiving copies of the data after each significant step [2].

Detailed metadata about the creation and location of each file must accompany every completed asset. Properly maintained metadata allow data managers to repair broken links within shape or image files and migrate data as operating systems or proprietary claims change. Therefore, preserving big data involves safeguarding all files generated by 3D-scanning technologies and all variations created during the process.

The open access movement in academia and data publishing is shaping the future of digital data curation and reuse, with data preservation playing a crucial role in this shift. As the volume of digital data increases, the solutions for their preservation are often misaligned with current workflows and proprietary interests, putting the digital record at risk of loss. To ensure the long-term viability of digital data, scholars must address these challenges.

**1. Understanding Technological Obsolescence**

Technological obsolescence occurs when a technology or digital format, once prevalent, becomes outdated due to the advent of newer, more efficient technologies. This phenomenon is not merely a matter of hardware but also encompasses software and digital file formats. For instance, a digital document saved in a proprietary format two decades ago may now be unreadable with present-time software. This issue is amplified in the context of cultural heritage data, which encompasses a vast array of formats, including text, images, audio, and video, each susceptible to obsolescence.

Solutions are being developed to ensure digital continuity, enabling vital data to be used in its original form and remain available for future use. These solutions are crucial for maintaining the integrity and accessibility of data over time, despite the fast-paced evolution of technology.

The lifecycle of digital 3D content in cultural heritage preservation relies on various information technologies and methodologies, integrating several key elements to ensure the creation, storage, management, processing, and access of digital content over time. These elements include:

- **Software packages** used for creating, storing, managing, processing, and importantly, providing access to digital content.
- Various **file formats** supported by different software at the time of creation, and the formats to which digital content is converted or transferred over time. This is due to the continuous evolution of software packages and file formats with updates, extensions, new trends, and features.
- The **digital storage media** where the content is initially stored and the media to which it is copied or transferred as time progresses.
- A mix of **operating systems**, **computer programs**, **security mechanisms**, **computer hardware**, and **communication networks**. These components support and enable the creation, management, protection, and crucially, the access to digital content over time.
- The evolving **standards** for formats and practices in digital preservation and information technology, which develop as new information technologies become more widely used and stable. These standards are crucial for communities responsible for digital content.

As the field of digital heritage continues to evolve, new technologies are emerging that have the potential to significantly impact the long-term preservation of 3D data. Technologies such as blockchain for immutable data records, machine learning for predictive data curation, and advanced data compression algorithms are increasingly being explored to address the challenges of technological obsolescence.

## 2. Digital Continuity, Data Management Plan and FAIR principles

## 2.2. Digital continuity

Digital continuity [3] is a concept that emphasises the maintenance and usability of digital information over time, despite changes in digital technology. It ensures that digital information remains complete, accessible, and usable, with an emphasis on managing information risks and technical environments. This includes strategies as file format conversion and information management, particularly crucial for organisations with a duty to maintain accountability and transparency, such as government and those in charge of infrastructure management. Moreover, digital continuity is vital for institutions like archives and libraries responsible for maintaining digital information repositories over time.

One approach to managing digital continuity, as practised by The National Archives in the UK, involves a four-stage process [4]:

- planning for action,
- defining digital continuity requirements,
- assessing and managing risks to digital continuity,
- and maintaining digital continuity.

This process embeds the concept of digital continuity into the digitization workflow, ensuring that organisations can continue to use digital information in the future despite technological changes.

## 2.3. DMP

In preserving 3D data for cultural heritage over the long term, it's essential to develop a comprehensive data management plan. This plan should address the entire lifecycle and long-term viability of the 3D data, considering factors like the type and volume of data being gathered and choosing file formats that promote both durability and broad compatibility. The plan must include robust and secure storage solutions, active throughout the project's lifespan and beyond, coupled with explicit protocols for data access, backup, and recovery to prevent data loss or corruption. Additionally, ensuring that data are accessible for future reuse is not only a hallmark of effective data management but also contributes to the project's visibility, and bolsters the transparency and reproducibility of its outcomes [5]. The FAIR principles provide guidelines designed to enhance the reusability of data, targeting both human users and machine processes [6]. In an era where data volumes are constantly expanding, it's impractical to concentrate solely on making data readable for humans. Consider, for instance, how indispensable search engines and web page search features have become in navigating this vast data landscape.

A Data Management Plan (DMP) should be created as a dynamic, guiding document, outlining the strategies for handling 3D cultural heritage data both during and after the project. This plan is not just a formal requirement but a blueprint for ensuring that the data remains valuable, accessible, and usable for the long term, contributing to the preservation and understanding of our cultural heritage [7].

## 2.4. FAIR principles

FAIR stands for Findability, Accessibility, Interoperability, and Reusability. A summary by Brinkman et al. [8], with some modifications and emphasis added, is as follows:

- **Findability**: This principle focuses on making data discoverable by both humans and computer systems. Achieving this involves the detailed description and indexing of data and metadata. Key practices include assigning persistent identifiers (PIDs) like Digital Object Identifiers (DOIs) for publications or ORCID iDs for individuals, citing research data, providing rich metadata following established schemas, incorporating keywords, and implementing dataset versioning.
- **Accessibility**: This pertains to the methods used for data access, which may include steps for authentication and authorization. It's important that metadata remains accessible even if the data itself becomes unavailable, allowing for the tracking of people, institutions, or publications. The use of open, freely available, and universally implementable access protocols, such as HTTP(S), and standardised exchange protocols like SWORD or OAI-PMH, is recommended.
- **Interoperability** involves the data (or metadata) being capable of integration with other datasets and systems. This means the data should seamlessly function with various applications and workflows used for analysis, storage, and processing. To achieve this, it is important to utilise formal, widely accepted languages, adhere to standard metadata schemas and vocabularies, and use qualified references. Employing open and commonly preferred file formats also enhances interoperability.
- **Reusability** focuses on maximising the potential for data reuse by clearly defining usage licences. Without a specified licence, the terms under which data can be reused remain unclear. Metadata should accurately reflect the data provenance, such as the author(s) and organisation(s), to establish its reliability and quality, and provide clear points of contact for any inquiries regarding reuse. Additionally, ensuring a clear understanding of the data is crucial for its reuse. This can be facilitated by providing codebooks, adhering to naming conventions, and offering other necessary explanatory material.

It's crucial to understand that 'FAIR' and 'open' are not synonymous. While open data sharing significantly enhances reusability, data can still adhere to the FAIR principles - Findability, Accessibility, Interoperability, and Reusability - even when they are not open.

This applies to data that cannot be openly shared, such as personal, sensitive, commercial, or security-related information. By providing detailed metadata, the existence and nature of such data can be made transparent, adhering to the principle of being "as open as possible, as restricted as necessary [9].

Implementing FAIR principles extends beyond merely making data open. It concerns ensuring that all data, including that which is restricted, is discoverable and usable within its specific constraints. The objective is to strike a balance, making data as accessible and usable as possible while respecting necessary limitations.

## 3. Public Repositories

The analysis identifies persistent challenges in expanding the scope of publicly hosted 3D model repositories. Despite ongoing efforts at the national level, issues such as the limited availability of models in public repositories and the preservation challenges posed by exceptionally large datasets remain prevalent. Furthermore, the dominance of specific platforms in the market presents a unique set of challenges that demand attention and strategic solutions [9] (for the list of repositories in the CH field see Appendix p. 29). While Sketchfab functions as a private viewer platform rather than a preservation repository, it remains the primary host for the majority of publicly accessible 3D models. The dynamics of platform businesses contribute to a trend where a single platform tends to dominate the market.

When considering the long-term preservation of data, particularly in the context of FAIR principles (Findability, Accessibility, Interoperability, and Reusability), it is crucial to choose the right repository. The primary goal is to ensure the data is stored for long-term use, but there are several key factors to consider in selecting a suitable repository:

- **Rich and Indexed Metadata**: The repository should support rich metadata, which is crucial for data discovery. Metadata should be indexed and utilise community-accepted controlled vocabularies, enhancing the discoverability and context of the data.
- **Open Protocols and Standards**: Utilisation of open protocols and standards is important for ensuring accessibility and interoperability of the data across different platforms and systems.
- **Persistent Identifiers (PIDs)**: Assigning PIDs to each dataset is essential. PIDs like DOIs (Digital Object Identifiers) ensure that datasets can be reliably cited, tracked, and retrieved, even as they are updated or moved.
- **Data Curation**: The repository should engage in active data curation. This includes monitoring and updating file formats to ensure long-term accessibility, as technologies and standards evolve.

Rich metadata not only aids in data discovery but also links to related information such as other datasets and publications. Even if the data itself cannot be published due to various reasons, the metadata should remain accessible. PIDs facilitate data discovery and support data versioning, search, citation, and retrieval. Clearly defined data access conditions and usage licences, available in a machine-readable format, are also crucial.

When evaluating a repository, it's important to check if it meets these criteria. A good indicator of a repository's compliance with long-term preservation standards is its certification. **Trustworthy repositories**[1], often endorsed with certifications like CoreTrustSeal, nestor-Seal DIN 313644, or ISO 16363[2], ensure the integrity and preservation of data. These certifications are markers of a repository commitment to maintaining high standards in data preservation. For repository operators, acquiring at least the CoreTrustSeal core certification is recommended, or working towards it.

In summary, selecting a FAIR-compliant repository is a key step in ensuring the long-term preservation and accessibility of your data. This involves considering factors like metadata quality, adherence to open standards, use of PIDs, and active data curation, along with verifying the repository's certifications.

### 4. Metadata and the Long-term Preservation

In the realm of long-term digital preservation, metadata are crucial for ensuring the longevity and ongoing accessibility of resources in the future and plays several critical roles. They aid in the unique identification and accessibility of digital resources, ensure their authenticity and integrity, manage legal rights and compliance, support preservation strategies, and provides context for interpretation and understanding. Metadata serves as the cornerstone for ensuring the long-term accessibility and usability of digital assets, particularly those complex as 3D data. Metadata facilitates the unique identification of digital objects, vital for locating and accessing specific resources in extensive digital archives. Functioning as a "digital memory", metadata encapsulates critical details such as provenance, change history, access conditions, and file formats, which are imperative for maintaining the integrity and interpretability of 3D objects over extended periods [10]. Furthermore, metadata is integral to the data migration process, facilitating the transfer of 3D models across various formats and platforms while ensuring their integrity and authenticity are maintained over time.

Regarding the support for the preservation of 3D data, metadata assume a unique role, since they can be used to document not only the technical specifications, such as dimensions and formats, but also intricate details like geometry, texture, and physical properties of the represented objects. This level of detail is crucial for future applications, including augmented reality or virtual reconstructions, where a thorough understanding of the structure and properties of the 3D objects is required and a clear distinction between the objects represented and the digital models that represent them is of primary importance. The role of semantic instruments, like ontologies, and other tools capable of unambiguously defining the identity of the objects described, such as persistent identifiers and shared vocabularies, becomes relevant in this sense.

---

[1] Also the term 'TDR' or 'Trustworthy Digital Repository' is used.
[2] "The CoreTrustSeal certification is envisioned as the first step in a global framework for repository certification which includes the extended level certification (nestor-Seal DIN 31644) and the formal level certification (ISO 16363)" https://www.coretrustseal.org/about/

In the context of the use of metadata to foster long-term preservation of digital information, several key initiatives stand aim at ensuring that digital information remains usable and understandable in the future, despite technological and environmental changes. One notable example among these is PREMIS [11] (PREservation Metadata: Implementation Strategies), a standard aimed at the long-term preservation of digital resources, widely adopted in libraries and archives to document the lifecycle of digital objects, but easily adaptable to the 3D context. PREMIS details metadata such as the identity, condition, context, and history of a digital object, including information about its creation, alteration, preservation, and access over time.

OAIS [12], the Open Archival Information System (OAIS) is another important framework widely recognized in the field of digital preservation. It is an ISO standard developed as a reference model for the archiving and long-term preservation of digital information. Its importance lies in providing a comprehensive, systematic approach to preserving digital information, ensuring it remains accessible and understandable over extended periods, by formulating policies for long-term digital preservation, providing a standard framework and best practices that guide institutions in developing robust preservation strategies. This model emphasises the importance of managing risks such as format obsolescence and data loss, while also highlighting the critical role of comprehensive metadata management. OAIS provides policies aimed at guiding institutions in developing robust preservation strategies to manage risks such as format obsolescence and data loss, while also highlighting the critical role of comprehensive metadata management.

Both of these initiatives highlight the importance, for quality preservation purposes, of generating metadata that documents the entire life of 3D models, encompassing the creation, management, archiving, publication, and ultimately the access and repurposing of 3D content. Furthermore, metadata should cover each of the defining aspects of digital objects, including technical details, descriptive data, provenance information, and so on.

**4.1. Types of metadata**

Metadata in archival systems typically fall into different categories, which however are not intended as rigid divisions, as they can share similar information. Each of these categories, if correctly structured, is able to provide a fundamental contribution to 3D long-term preservation policies.

**Descriptive metadata** support the identification and discovery of 3D content since they describe a resource for purposes of identification, discovery, and selection. They can include elements such as title, author, abstract, keywords, and resource identifiers. They also provide information about the context in which a digital object was created and used, crucial for future interpretation. Standardised and well-documented descriptive metadata promote interoperability between different systems and archives, facilitating global exchange and sharing of digital resources. Descriptive metadata are becoming increasingly important because they can be used to describe, organise, and package the digital object's files. Since digital objects are not self-describing, descriptive metadata are paramount for identifying semantic-level content and provide context. Furthermore, without this information, verifying whether an object is the original, a copy, or a fabricated or fraudulent item is impossible in most cases [13]. Currently, Dublin [14] Core and METS [15] are the most used standards for this type of information, although other models such as Schema.org are gradually becoming popular, especially in relation to the context of the semantic web.

**Structural metadata** are used to organises files into coherent objects since they provide information about the internal structure of a digital resource, like how pages are organised in a book or chapters in a multimedia document or, in the case of 3D, how its various digital parts organised and coordinated, and thus are useful for navigating within the resource. METS and MODS are the most popular models for encoding this type of metadata.

**Administrative metadata** relate to the management and administration of the resource. They can include information about digital objects' creation, copyright, licensing, access restrictions, and preservation. Administrative metadata help users to understand how they can legally access and use digital resources, and institutions to manage compliance with privacy and data preservation laws. They are often used to manage digital resources over time and establish object quality. Administrative metadata can also include checksums or other integrity information to ensure that a digital object has not been altered or corrupted over time, and can help in planning for data migration, emulation, and other strategies to keep digital objects accessible over time despite technological obsolescence. The standard proposed by PREMIS is particularly suitable for encoding administrative metadata necessary for long-term digital preservation.

**Provenance metadata**, also known as **paradata**, are used to document the history of a digital object, including details about its creation and modification, and what processes or changes it has undergone. This type of metadata is very important for understanding the context of a resource and establishing its authenticity and integrity. The PROV [16] model, developed by the W3C consortium, offers an excellent example of how to encode this type of information.

**Technical metadata** describes the technical characteristics of digital objects, such as file format, resolution, and system requirements, which are vital for long-term preservation planning. Under certain conditions, technical metadata can be extracted from 3D models through specific procedures involving specialised software. For example, it is possible to derive information about interactive 3D objects and scenes from the VRML format, facilitating their distribution and visualisation on different platforms. The same can be done with formats such as IFC, natively designed in the context of BIM as a rich set of metadata to describe buildings and their components in detail. Technical metadata of 3D models typically includes detailed information such as dimensions, scale, material properties, and geometric structure. This ensures that critical technical aspects of the 3D models are documented and preserved for future reference and use.

**Preservation metadata** is a particular category of metadata, especially defined within PREMIS to support digital preservation by maintaining authenticity, identity, renderability, understandability, and viability of the digital resources. Preservation metadata are not bound to any one category as they comprise multiple types of metadata. The importance of preservation metadata is not merely about the physical survival of data but more crucially about preserving its comprehensibility and relevance, an aspect becomes particularly significant in disciplines such as digital archaeology, cultural heritage conservation, and scientific research, where the fidelity and authenticity of data over time are paramount. Preservation metadata acts as a bridge across temporal gaps, ensuring that future generations are equipped with the necessary context and understanding of the original use and significance of the 3D objects.

### 4.2. Building valuable metadata

The definition of high-quality preservation metadata for 3D digital content requires the use of tools like Persistent Identifiers (PIDs), controlled vocabularies, and ontologies to ensure the persistent accessibility and identification of 3D objects, the standardisation and clarity of the information, and the implementation of structured frameworks for representing complex relationships and concepts within 3D content, thereby enhancing the metadata's depth and contextual relevance. The synergy of these tools elevates the quality of metadata, and guarantees a comprehensive, accurate, and sustainable documentation of 3D digital assets for long-term preservation and accessibility.

**Persistent identifiers**

The integration of persistent identifiers such as Digital Object Identifiers (DOI) and Archival Resource Keys (ARK [17]) into the metadata of digital resources is a strategic measure for bolstering long-term preservation efforts. In metadata, these identifiers serve as immutable reference points that facilitate the tracking and retrieval of digital objects amidst the dynamic landscape of digital storage and content management. They enable a stable, persistent method of access and citation, essential for scholarly communication, digital archiving, and data integrity. When embedded in metadata, DOIs and ARKs ensure that each digital object can be uniquely distinguished and persistently located, thereby enhancing the metadata's role as a carrier of essential information across the temporal continuum of digital preservation. Their use in metadata is critical not only for maintaining the link between data and its descriptive information but also for supporting robust archiving strategies that accommodate future technological shifts and migration processes.

**Controlled Vocabularies**

To implement optimal long-term preservation policies, it is necessary to ensure that metadata is structured in such a way as to be easily accessible and interoperable with other systems and platforms, facilitating the retrieving, sharing and use of 3D models in different scenarios. Both PREMIS and OAIS emphasise the use of controlled vocabularies, perceived as fundamental tools, particularly in the management of metadata content and in enhancing the accessibility and comprehension of stored information. Controlled vocabularies aid in standardising metadata terminology, providing consistency and accuracy in the description of archived information. Thus, they are vital for achieving interoperability among different archival systems, offering a common language that aids in information sharing and exchange. Controlled vocabularies also contribute to preserving the context and meaning of the archived information, ensuring that it remains understandable over time.

**Ontologies**

Ontologies further enrich the landscape of the metadata models and standards described above by providing structured tools that define types, properties, and interrelationships among concepts in a particular domain. Ontologies as the CIDOC CRM, CRMdig and others from the family of CRM-compatible models enable creating high-quality metadata for digital objects, thanks to their ability to clearly define and organise the concepts and relationships within a domain, thus ensuring the consistency and accuracy of the encoded information, which is essential for effective digital preservation. They are instrumental in standardising concepts and relationships, thereby enhancing interoperability and data sharing across various systems and organisations. This standardisation is crucial for maintaining the integrity and utility of metadata over time, especially in complex and evolving fields like digital preservation. Furthermore, ontologies enable the creation of rich, detailed representations of digital objects and significantly improve search capabilities through their complex structures. This detailed representation aids in capturing the nuances and complexities of digital objects and guarantees a comprehensive and multifaceted approach to long-term digital preservation.

Ontologies also play a fundamental role for "semantic preservation", involving the long-term understandability of metadata, as outlined by OAIS to maintain the comprehension and interpretation of digital information and the related metadata over time. For example, in preserving a dataset of scientific observations, semantic preservation involves more than just retaining raw data. It includes preserving the context of how and why the data were collected, describing it in a format capable of remaining clear and unambiguous over time. This ensures that future researchers can understand the data's original context and significance, and the complex relationships and meanings inherent in the data, enabling future users to interpret the information accurately as intended in its original creation. Such approach also ensures that the data remains not only accessible but also meaningful and usable for future generations.

### 4.3. Managing metadata and paradata

Starting from the 2006 London Charter [18] that gave the first guidance on the topic, one of the first publications dedicated to directions in Virtual Heritage is offered by Bentkowska-Kafel et al. [19]. The publication emphasises the importance of viewing three-dimensional visualisation as both a valuable intellectual endeavour and a legitimate methodology for historical research and communication. It advocates for intellectual clarity in research that relies on visualisation, examining this concept from various disciplinary theoretical and practical standpoints. The recommendation is to accurately record paradata alongside specific research findings, treating it as a fundamental element of virtual models. It is additionally advised to preserve this paradata beyond the lifespan of the technology used for visualisation.

Considering that during the period in question CH repositories started to grow and 3D data started to be included, the publication by Bentkowska-Kafel et al., together with the one by D'Andrea & Fernie (2013) [20], gave support to the entire process. D'Andrea & Fernie suggested a metadata schema specifically for 3D cultural objects, aimed at enhancing the description and management of CH repositories that include 3D items. This schema, drawing upon the foundations laid by the CARARE project, emphasises a clear approach to detailing the characteristics of cultural objects, the digitization techniques and methodologies used, and the rationale behind the creation of digital objects. Their paper delves into the provenance aspect within the CRMdig schema and the paradata principles of the London Charter. It also explores how provenance and paradata could align with Europeana's strategy to improve the efficiency of resource reuse and enhance the usability of online resources. In line with this, Europeana has continuously worked to develop and improve its Europeana Data Model (EDM). This model focuses on collecting, connecting, and enriching the descriptions from its content providers, while also addressing the multifaceted data quality challenge, particularly in terms of the reuse and discovery of cultural heritage objects.

The 2020 study by Blundell and colleagues [21] focuses on the specific needs for metadata in 3D data. It outlines steps in the digital asset lifecycle, such as creation, management, publishing, accessing/reusing, and archiving, and suggests appropriate metadata fields for each step. The study also anticipates future metadata requirements like annotations and metadata that enhance dataset accessibility, discovery, and usage. Additionally, it discusses the importance of data quality and its relevance for reuse and reproducibility.

Highlighting the increasing number of publications on this subject, the paper references another work by Huvila in 2022 [22]. Huvila's research emphasises the importance of understanding and documenting paradata, which refer to the background information regarding the creation, management, and usage of research data. This is crucial for making data meaningful and useful for future users. The article points out the challenges in capturing the paradata and stresses the need to consider the diverse users and applications of data. It argues against the assumption that descriptions of processes remain consistent over time and place, advocating for a more dynamic and user-oriented approach.

In a similar effort to create rules for preserving 3D data, a study by Moore and colleagues in 2022 emphasises the importance of developing common guidelines, procedures, and standards. This research highlights the best methods for preserving, managing, and providing access to 3D data, as well as dealing with metadata and legal aspects. The authors also offer suggestions for adopting these standards and point out areas that need more work.

For over twenty years, researchers in cultural heritage and archaeology have been working to develop essential metadata and paradata for 3D data across various stages of their projects. This effort has recently resulted in two significant 2022 publications that outline the necessary and suggested metadata for 3D datasets (as detailed by Medici and Fernie [23], and Moore et al [24]). However, the challenge lies in achieving broad acceptance and integration of these standards. Furthermore, there are ongoing issues regarding the best methods and locations for recording and disseminating this metadata.

## 5. Metadata Schema

For effective collaboration among various projects and repositories, adopting a standard metadata schema is recommended. General metadata schemas cater to a wide range of datasets, encompassing fundamental elements like 'author(s)', 'date', 'language', and 'description', with Dublin Core being a prominent example. On the other hand, domain-specific metadata schemas are tailored to particular fields or professions, such as cultural heritage, libraries, museums, and archives. These specialised schemas blend general elements with more detailed resource descriptions. Notably, there's no universal standard covering all cultural heritage aspects. Key schemas in this domain include CIDOC CRM, CRMdig, LIDO, Smithsonian, CARARE, and Europeana EDM, as outlined in Table 1.

*Table 1 Recommended metadata schemas for cultural heritage use (source:'4CH - D4.1 Report on Standards, Procedures and Protocols [23]').*

| Name | Characteristics, themes, focus | Link |
| --- | --- | --- |
| ARCO | Museums, 3D models of museum artefacts. | |
| CARARE 2.0 | Monuments, buildings, landscape areas; 2D and 3D. Application profile of MIDAS with extensions to support the EDM and the CIDOC CRMdig. | https://pro.carare.eu/en/introduction-carare-aggregation-services/carare-metadata-schema/ |
| COSCH | Spatial and spectral recording of material cultural heritage. | https://link.springer.com/chapter/10.1007/978-3-319-75789-6_5 |
| CRMdig | Provenance of digital objects; 2D and 3D. Extension of CIDOC CRM. | https://cidoc-crm.org/crmdig/ |
| Europeana Data Model (EDM) | Metadata from museums, libraries, archives, galleries (GLAM); various types of digital models including 3D (but limited coverage?) | https://pro.europeana.eu/page/edm-documentation |
| INCEPTION H-BIM | BIM model and CH information, architectural; 3D. | https://link.springer.com/chapter/10.1007/978-3-030-12960-6_23 |
| LIDO | Museums, museum objects | https://cidoc.mini.icom.museum/working-groups/lido/lido-overview/about-lido/what-is-lido/ |
| Mainzed | 3D capturing processes | https://heritagesciencejournal.springeropen.com/articles/10.1186/s40494-021-00561-w#Sec31 |
| METS | Digital libraries | https://www.loc.gov/standards/mets/ |
| PARTHENOS | extension of CIDOC CRM and CRMdig for research infrastructure aggregators. | https://cidoc-crm.org/Resources/parthenos-entities-research-infrastructure-model |
| Smithsonian 3D metadata model | Museum 3D programmes | https://dpo.si.edu/blog/smithsonian-3d-metadata-model |
| STARC | 2D and 3D. CARARE 1.0, CRMdig | |

## 6. Serving Diverse User Communities

The versatility of three-dimensional heritage objects is explored, spanning diverse disciplinary contexts such as art and architectural history studies, museology, archaeology, and heritage conservation [26,27].

User-friendly interfaces and workflows are pivotal elements in encouraging content provision and fostering a collaborative environment. Modern virtual reality (VR) technology has garnered significant attention for its ability to immerse users within geospatial data sets [28].

Managing large datasets with version controls, customising labelling workflows, enhancing annotation precisions, and gaining full visibility of datasets with automated tools and advanced search for both 2D and 3D visual data are also important features.

## 7. Feature and Quality Requirements

The primary obstacle is the lack of consistency among 3D models. Currently, 3D models are often created independently, using diverse base (sensor) data, reconstruction techniques, and software. This results in significant variations in geometry (e.g., surface collections versus volumetric representations), appearance, and semantics. Additionally, models are stored in different formats (such as XML, graphics, or binary formats), leading to variations in their underlying data models. Even models initially identical can diverge due to independent processing, either through mismatched updates or conversions between different formats aimed at resolving software incompatibilities. These discrepancies have practical implications, influencing the applications for which a 3D model can be utilised, the required processing steps, and the likelihood of errors in the final output. Therefore, it is crucial to understand how 3D models are constructed and explicitly document this information in the model's metadata.

As underscored by Biljecki et al. [29], many openly available 3D models exhibit numerous geometric and topological errors, ranging from duplicate vertices to missing surfaces and self-intersecting volumes.

A potential avenue to mitigate geometric errors involves automatic repair algorithms present. However, it's important to note that these algorithms often involve a semi-manual process, and there is a risk that rectifying one error may inadvertently introduce new issues elsewhere in the model. Despite these challenges, addressing and improving the quality of 3D models is essential to unlock their full potential as a shared and interoperable platform for different applications.

'Higher quality' does not necessarily mean 'greater precision'; it means up-to-date 3D data without errors and aligned with the specific needs of specific applications rather than serving visualisation purposes only.

### 7.1. Monitoring and Fostering Standards

In the realm of 3D models for cultural heritage, achieving standardisation is crucial to maintaining consistency in both geometry and semantics. Establishing and adhering to common standards ensures that these digital representations of cultural artefacts, monuments, and historical sites can be effectively shared, exchanged, and utilised across various platforms and applications.

One noteworthy standard in this context is the use of widely accepted formats and specifications such as glTF (Graphics Library Transmission Format), OBJ (Wavefront .obj file) and COLLADA. These formats provide a standardised way to represent 3D models, allowing for interoperability among different software applications and platforms.

Moreover, metadata standards play a significant role in documenting the construction and contextual information of 3D models in cultural heritage. Metadata standards can include detailed descriptions, historical context, and information about the digitization process. This not only contributes to the understanding of the cultural significance of the represented artefacts but also facilitates the proper use and interpretation of the 3D models.

In addition to file formats and metadata, the adoption of standardised practices in 3D scanning technologies is essential. Common methodologies, such as photogrammetry or laser scanning, can be employed with standardised parameters, ensuring consistency and accuracy in the acquired 3D data.

Efforts are also being made to develop domain-specific standards for cultural heritage 3D models. These standards may address the unique requirements of preserving and presenting cultural artefacts, architectural structures, or archaeological sites. Such domain-specific standards help capture the nuances of cultural heritage and enhance the overall quality and authenticity of 3D models within this context.

Standardisation is of paramount importance to ensure consistency in both geometry and semantics across 3D models.

The conversion of semantic 3D models, irrespective of format, poses challenges, encompassing both geometric considerations and issues related to incompatible semantics. In the context of the Industry Foundation Classes (IFC) standard commonly used in Building Information Modeling (BIM), the integration of highly detailed models from the design and construction phase into a 3D model for different uses presents complexities.

Historic Building Information Modeling (HBIM) [30] provides a detailed and structured approach to documenting the architectural and historical features of heritage sites. It integrates various data types, such as laser scans and historical records, into a single cohesive model. However, ensuring the semantic classification aligns with heritage-specific requirements remains a challenge. Current IFC standards, widely used in the AEC/FM (Architecture, Engineering, Construction/Facility Management) industry, face difficulties in fully accommodating the complex and specific needs of cultural heritage data. This includes representing historical changes, material degradation over time, and various layers of renovation and conservation efforts.

The IFC schema is implemented in BIM environments through Model View Definitions (MVDs) [31], which define subsets of the IFC schema necessary for specific data exchange requirements. The creation and implementation of MVDs are complex and require deep knowledge of the IFC schema. Furthermore, existing software must support these MVDs, which is not always the case, leading to potential interoperability issues. Researchers have proposed extending the IFC schema to include heritage-specific data by developing new property sets and classifications that can accurately capture the unique characteristics of heritage buildings. However, these extensions need to remain within the boundaries of the IFC schema to maintain interoperability [32].

Automating the conversion between IFC models proves non-trivial due to intricate mappings between semantic classes, where different semantic information is linked to geometrical primitives in the different standards.

The semantic disparities, coupled with variations in software and geometric modelling paradigms, make reusing data across different domains challenging.

## 8. Common 3D File Formats in Cultural Heritage Preservation

3D file formats contain data representing three-dimensional space and the necessary information for displaying these data. These formats have a wide range of uses across various fields, including:

- Preserving cultural heritage, by creating digital replicas of museum artefacts.
- Documenting, for restoration and conservation, buildings, archaeological sites, and historical structures.
- Creation of historically accurate scenes and objects for visual effects, video games, and animation.
- Designing 3D printed objects.
- Developing files and systems for Virtual Reality (VR) and Augmented Reality (AR).
- Creating digital artworks.

Application of certain file formats is preferred because they offer a better long-term guarantee in terms of usability, accessibility, and sustainability [33]. If file formats are frequently used, have open specifications, and are independent of specific software, developers, or vendors, the files are more easily accessed and reused and for a longer time. It is important to note though that in practice it is not always possible to meet all three criteria, and that a balance needs to be found. There may be options to convert a non-open file format to an open one - here, it is, however, important to assess if all information is retained. Metadata should always be available in an open format; however, when converting from an original proprietary format into an open format, information may be lost. Only keeping the proprietary format is also not ideal, as not everyone may have access to the software or it may become obsolete. If possible (in terms of costs, for example) one should keep both the original, proprietary format and the open format.

No single format is ideal for preserving and using 3D data in the future. The choice of file format should depend on the specific features and functions that need preserving and the intended future applications.

The file formats used in cultural heritage are also used in other disciplines, and recommendations of preferred file formats have been compiled, for example, by the Expert Group on Digital Cultural Heritage and Europeana [27], DANS and UKDS [35]. Preferred formats specific to cultural heritage types of data have also been compiled by, for example, Archeological Data Service (ADS) [36]. A summary of preferred formats for 3D data is presented in Table 2. Generally recommended are ASCII-based files for point coordinates and commonly used, ideally open file formats. Formats like Autodesk FBX (.fbx), Blender (.blend), or 3D PDF (.pdf) are in general deprecated for long-term preservation.[3]

Digital projects in Cultural Heritage (CH) involving 3D data are inherently multimodal, balancing various types of data depending on the project goal. A key aspect of any 3D workflow is that data undergo numerous changes and are processed through different software. Even projects focused solely on capturing an object geometry, while overlooking colour and texture, rely on temporary data like 2D images or point clouds. This shows that there is no distinct separation between 2D and 3D workflows. Preserving these different data types, even if it may seem irrelevant for the current project, can help address the issue of diversity in data types. It is also important to preserve the distinction among reality-based data, born-digital data, and processed reality-based data.

---

[3] For other purposes preferred file formats may differ, see for example the Unreal Engine guidelines: 'Supported 3D File Formats'. Available at: https://support.fab.com/s/article/Supported-3D-File-Formats (Accessed: 26 June 2024).

The complexity of 3D data formats goes beyond their structural specifics. Their role in how diverse software and platforms interpret data is critical for interoperability—a cornerstone for effective utilisation of 3D CH data. The European Commission's recent contribution in this area includes a comprehensive enumeration of current 3D formats, covering both raster and vector types [37]. This compilation not only serves as a resource but also highlights the concerted efforts of international standardisation bodies.

The extensive range of these formats and their adoption as standards across various industries underscore the significance of uniformity in 3D CH data handling. Key players in this standardisation process include the European Committee for Standardization [38], ISO (International Organization for Standardization), and the Web 3D Consortium [39].

*Table 2 Preferred formats for 3D data (different types) based on DANS,[4] ADS,[5] and inputs from the community at the workshop 'Shaping the World of 3D'. See also UKDS for recommended and other acceptable formats in general (but no recommendations are included for 3D file formats).[6] From 4CH 5.1 Deliverable.*

| Format | Extension | Included list | Remarks |
|---|---|---|---|
| **WaveFront Object** | .obj | DANS; ADS; EU Digital Strategy[1] | ADS: for wireframed or textured models |
| **Polygon file format** | .ply | DANS; EU Digital Strategy | ADS: ASCII version suitable if file content is clearly documented |
| **X3D** | .x3d | DANS, ADS, EU Digital Strategy | ADS: recommended for complex 3D content |
| **COLLADA** | .dae | DANS; EU recommendation | ADS: recommended where X3D is not an option |
| **Standard Tesselation Language** | .stl | ADS | ADS: ASCII format suitable for very basic datasets. Popular in 3D printing and computer-aided manufacturing. |
| **Virtual Reality Modelling Language** | .wrl, .vrml, .wrz | ADS | ADS: now replaced by X3D |
| **Autodesk Drawing Interchange Format** | .dxf | ADS | ADS: only suitable for preservation of native CAD datasets |
| **glTF** | .gltf; .glb | 4CH workshop[1], EU Digital Strategy | Designed for efficient transmission and loading of 3D models in applications. |
| **Draco** | .drc | 4CH workshop | Draco is a compression library for 3D geometric meshes and point clouds. |
| **LASer** | .laz /.las | 4CH workshop | These formats are particularly important in the field of cultural heritage for recording and analysing natural and man-made landscapes |
| **E57** | .e57 | EU Digital Strategy | The E57 file format is specifically designed for storing point cloud data. |
| **Industry Foundation Classes** | .ifc | 4CH workshop; EU Digital Strategy | IFC is a standardised, open file format used primarily in the building and construction industry for Building Information Modeling (BIM). |

---

[4] 'File formats' *DANS*. Available at: https://dans.knaw.nl/en/file-formats/ (Accessed: 26 June 2024).
[5] 'File formats' *ADS - Archaeology Data Service*. Available at: https://archaeologydataservice.ac.uk/help-guidance/guides-to-good-practice/data-analysis-and-visualisation/3d-models/creating-3d-data/file-formats/ (Accessed: 26 June 2024).
[6] 'Recommended formats' *UK Data Service*. Available at: https://ukdataservice.ac.uk/learning-hub/research-data-management/format-your-data/recommended-formats/ (Accessed: 26 June 2024).

**Conclusions**

The preservation of 3D cultural heritage data requires a comprehensive approach that integrates advanced technological strategies and accurate management practices, recognizing the inherently multimodal nature of digital heritage projects, which often combine various data types tailored to specific objectives. Throughout the workflow, data undergo frequent modifications and are processed using different software applications, which can sometimes lead to information loss. For example, projects focused primarily on capturing an object geometry might overlook aspects such as colour and texture, relying instead on temporary data like 2D images or point clouds. To mitigate this issue, it is crucial to preserve such diverse data types. Although it might seem irrelevant at the moment, they play a crucial role in addressing data diversity challenges and ensuring the completeness of digital heritage documentation.

Current institutional recommendations and European Union reports suggest adopting formats like OBJ, PLY, and COLLADA for their accessibility and lightweight nature, and glTF for advanced material representations like Physically Based Rendering (PBR). While these formats sometimes compete to fully support complex representations, they are generally accompanied by comprehensive documentation — including raw files, linked files, associated data, metadata, and paradata — which significantly reduces their limitations.

Challenges associated with the 3D digitization of cultural heritage include the longevity of software and hardware, but the noticeable underutilization of open-source software applications is particularly striking. The guiding principles for digital preservation emphasise maximising the use of open solutions to ensure better long-term access and preservation.

HBIM provides a detailed and structured approach to documenting the architectural and historical features of heritage sites. It integrates various data types, such as laser scans and historical records, into a single cohesive model. However, ensuring the semantic classification aligns with heritage-specific requirements remains a challenge. Current IFC (Industry Foundation Classes) standards, widely used in the AEC/FM industry, face difficulties in fully accommodating the complex and specific needs of cultural heritage data. This includes representing historical changes, material degradation over time, and various layers of renovation and conservation efforts. Researchers have proposed extending in various ways the IFC schema to include heritage-specific data by developing new property sets and classifications that can accurately capture the unique characteristics of heritage buildings. However, these extensions need to remain within the boundaries of the IFC schema to maintain interoperability.

The choice of appropriate formats becomes particularly important when considering complex 3D scenes, such as those used in virtual reality animations that might include not just geometry, but also materials, textures, lighting, cameras, viewpoints, character animations, physical interactions (like collisions and gravity), and sounds. Preserving all these elements in their original formats is required to fully appreciate and reuse them. Formats like glTF offer a balanced approach, navigating the compromise between interoperability and quality of representation while ensuring the original data is maintained.

Looking ahead, the future development of 3D formats will require careful observation to determine which solutions are best to adopt. A balanced approach in selecting file formats is essential, weighing the trade-off between interoperability and the quality of representation, and always ensuring that the original data is preserved. This comprehensive and adaptive strategy is crucial for ensuring that 3D cultural heritage data remains accessible and interpretable for future generations, while retaining the rich cultural significance they represent.

## 9. Case Study: The *Agios Ioannis Lampadistis* Monastery, Cyprus

### 9.1 Methodological Workflow for Preservation

The Cyprus Institute has developed a comprehensive methodological workflow tailored to the preservation needs of the *Agios Ioannis Lampadistis* Monastery, This work, published as part of the 4CH project [43] and currently under development, is designed to address various aspects of preservation, from initial assessment and documentation to continuous monitoring and intervention.

#### 9.1.1 Initial Assessment and Documentation

The preservation process begins with a thorough assessment of the site. This involves evaluating the asset's historical context, architectural features, and cultural significance. A comprehensive review of historical documents, architectural plans, and previous research is essential, as well as conducting an accurate inspection to determine the current state of preservation, identifying areas of damage, decay, or alteration. It is also required to assess potential risks, including environmental factors (e.g., weather, pollution), human activities (e.g., overtourism, vandalism), and natural disasters (e.g., earthquakes, floods).

This initial phase is required to understand the present condition of the site and forms the foundation for all subsequent preservation efforts. It is also vital for planning and organising the surveying and acquisition work. The information gathered from these investigations helps to identify which areas require detailed and precise documentation, either due to their fragility and susceptibility to deterioration or because of their cultural importance.

Historical and recent documentation, as well as surveying work conducted for maintenance and restoration, are of paramount importance. These records capture any modifications to the building, thereby enabling the recovery of critical information. One of the most intriguing aspects is the lack of standardisation in past documentation, whether due to the use of proprietary software or a general lack of interest in documenting every aspect of the work.

Therefore, it is essential to recover and convert all materials into a more accessible format, adhering to the FAIR principles (Findability, Accessibility, Interoperability, and Reusability), before proceeding to the next phase.

For the Lampadistis Monastery, both historical and digital documentation efforts are complemented by comprehensive historical studies and architectural analyses. These studies are crucial for understanding the evolution of the monastery's structures and evaluating the impact of previous conservation efforts.

Advanced technologies such as Terrestrial Laser Scanning and UAV (Unmanned Aerial Vehicles) platforms have been employed to create high-resolution 3D models of the monastery. Such 3D models facilitate a thorough analysis of the monastery's condition, aiding in effective conservation planning and implementation.

#### 9.1.2 Data Curation and HBIM Integration

HBIM is employed to curate and manage the extensive data gathered during the documentation phase. HBIM serves also as a digital repository and interactive tool, enabling stakeholders to access detailed information about the site's structural components, historical context, and conservation status.

**Mesh Models and Semantic Segmentation**: The creation of detailed mesh models and the segmentation of the monastery into semantic areas facilitate targeted analyses and conservation planning.

**Data Enrichment**: The HBIM model is enriched with various property sets, incorporating historical events, conservation states, and risk assessments. This comprehensive dataset supports informed decision-making and effective preservation strategies.

### 9.2 Collaborative Platforms and Public Engagement

To enhance accessibility and promote collaboration among stakeholders, the HBIM model is integrated into web-based platforms such as the Inception Core Engine (ICE) Viewer. This platform allows stakeholders to interact with the data, contribute to ongoing conservation efforts, and share insights [44].

#### 9.2.1 ICE Viewer

This open-standard Semantic Web platform enables the visualisation and interaction with the HBIM model, providing access to all stored information and facilitating collaborative analysis and decision-making (Fig. 1).

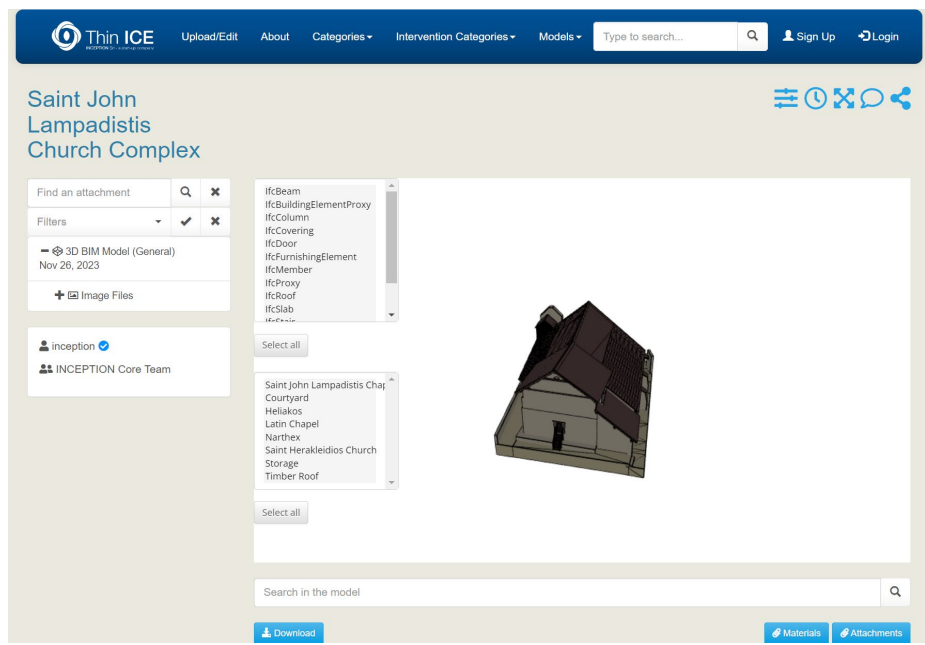

*Fig. 1.3D model of the Lampadistis Monastery created using BIM methodology within the Inception Core Engine (ICE) Viewer.*

The viewer allows users to associate each element with comprehensive scientific documentation such as X-ray Fluorescence (MA-XRF) data or historical images related to the selected elements. It is an excellent tool for providing context to analyses and documentation directly linked to the 3D model [43] (Fig.2).

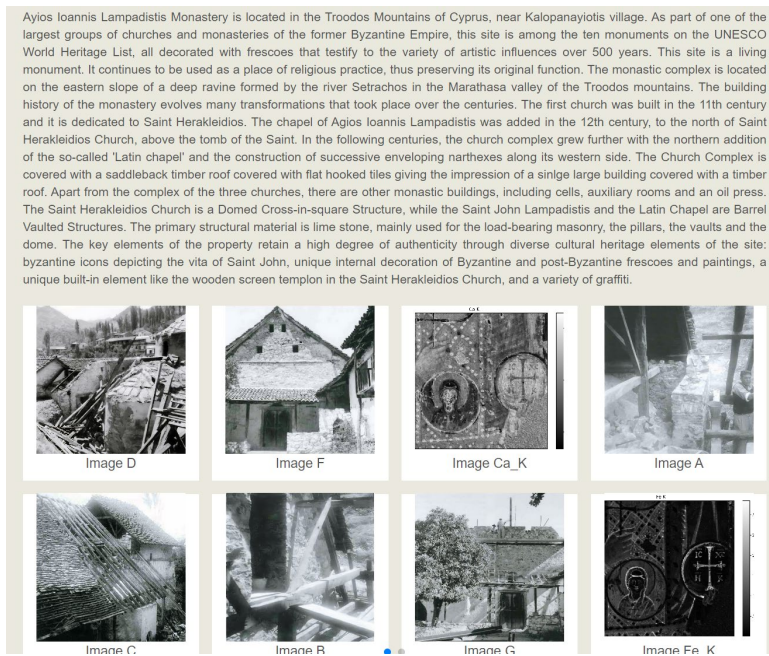

*Fig 2. scientific documentation; X-ray Fluorescence (MA-XRF) data and historical images within the Inception Core Engine (ICE) Viewer.*

Additionally, the ability to download all associated data aligns perfectly with the FAIR principles (Findability, Accessibility, Interoperability, and Reusability) and the concept of open data.

This functionality not only enhances the collaborative aspect of heritage conservation by enabling various stakeholders to access and contribute to a centralised data repository but also supports the integration of diverse data types, enriching the overall understanding and preservation strategy of cultural heritage sites.

While ICE Viewer and visualisation tools offer numerous benefits for enhancing the documentation, analysis, and engagement of cultural heritage sites, they also present several challenges. These include technical complexity, data management issues, potential interoperability problems, and high costs. Addressing these challenges requires careful planning.

**Visualisation Tools**: tools like the Potree viewer [46], 3DHOP [47] and 360° panoramic tours are implemented to engage the public and promote the site. These tools enhance the visitor experience and support educational and research activities by making the site's data widely accessible.

While the **Potree** viewer (Fig. 3) is a versatile tool for 3D visualisation, using models without colours introduces several critical limitations. These include reduced visual detail and realism, challenges in data interpretation, diminished user engagement, and potential technical constraints. For cultural heritage applications, where detail and accuracy are paramount, incorporating colour into point cloud models can significantly enhance the effectiveness of visualisation tools like Potree [48].

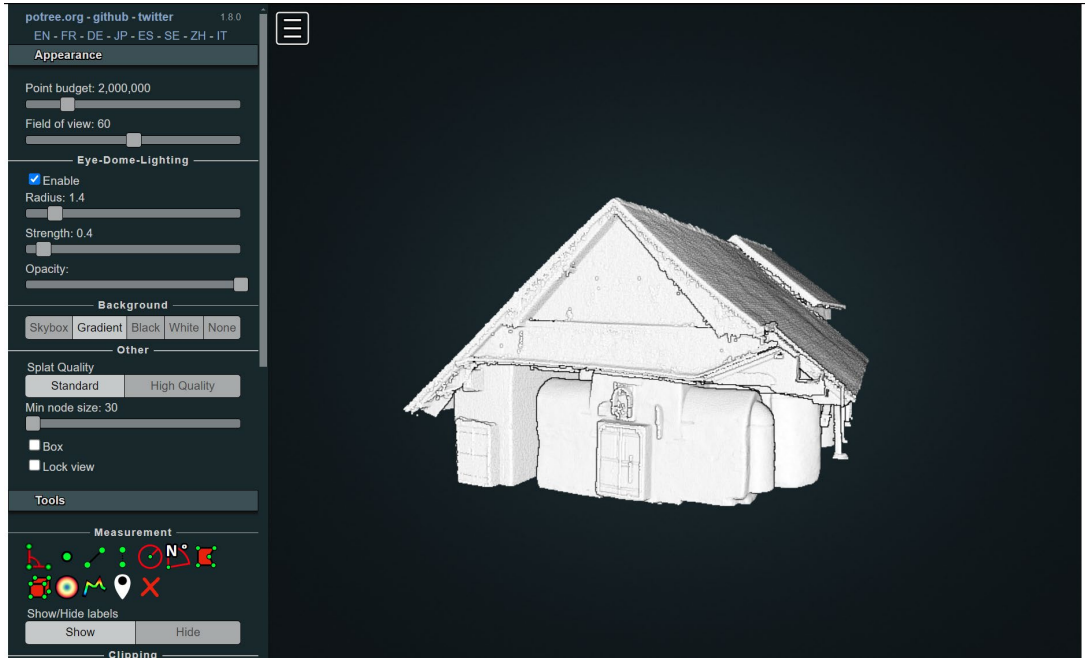

*Fig. 3. 3D point cloud of the Lampadistis Monastery within the Potree viewer*

**3DHOP** (3D Heritage Online Presenter) is an open-source framework designed for the creation of advanced web-based visual presentations of high-resolution 3D content, specifically catering to the needs of the Cultural Heritage field.

3DHOP offers a robust solution for the web-based visualisation of cultural heritage 3D models, providing high-resolution streaming, ease of use, and interactive features that enhance user engagement. However, challenges such as technical expertise requirements, performance limitations on older hardware, and the need for effective data management must be addressed to fully realise its potential.

In this context, researchers at CYI (The Cyprus Institute) are studying solutions for a viewer that can meet the research and visualisation needs. This research aims to overcome the existing challenges by developing a more accessible, high-performing, and efficient data management system, thus enhancing the overall experience and usability of 3DHOP for cultural heritage projects.

### 9.4 The Agios Ioannis Lampadistis Metadata

### 9.4.1 The STARC Metadata System

The case study of the Agios Ioannis Lampadistis Monastery deals with different types of metadata utilised to ensure its long-term sustainability. Different types of metadata have been created for each step of the digital information acquisition process, according to the most recent international recommendations and good practices.

For ensuring long-term preservation metadata, the Cyprus Institute natively encoded the information about the monastery using the STARC metadata system [49], a standard that provides a comprehensive framework for documenting the provenance, description, and technical specifications of digital resources.

Using the STARC schema, descriptive metadata are defined for general information about the Lampadistis monument, including its name, description, and location, a detailed description of its history, architecture, and significance. Administrative metadata provides information about the management and ownership of this cultural heritage site. This type of metadata is essential for ensuring long-term sustainability, as it helps to manage access rights, preservation policies, and data ownership. In the case of the Lampadistis Monastery, the administrative metadata includes information about the function of the monastery as a church and its status as a UNESCO World Heritage Site. Technical metadata provides information about the technical aspects of the cultural heritage site, including its construction materials, structural components, and construction techniques, essential for targeted analyses and conservation planning, as they help to identify areas of damage, decay, or alteration and supports effective preservation strategies. Technical metadata also include details about the monastery's architecture, such for instance, the domed cross-in-square structure of the main church, the chapel with the relics of Agios Ioannis Lampadistis, and the common narthex. Finally preservation metadata provides information about the preservation efforts and history of its development as a cultural heritage site. They also include information about the construction of the monastery and the renovation history, as well as its current state of preservation, and are fundamental to describe and preserve the site's historical context, its cultural importance and authenticity.

The STARC standard also provides a comprehensive collection of metadata for the 3D models and other digital objects related to the cultural complex. In this case, descriptive metadata carries information about the digitisation events and actors, essential for understanding the context of the digital resource and its cultural significance. In addition, STARC also provides technical metadata that describe the digitization process, its features, and detailed information about the resulting digital resource. This includes data about the equipment used, the digitization technique employed, and the resolution of the generated 3D models. STARC also allows the encoding of more detailed information about the digital resource.

For example, the 3D model of the church includes metadata about the data acquisition process, including the technology used (Structure-From-Motion), the software used (Metashape), the camera model and serial number, the lens model and aperture, the focal length, exposure time, number of cameras, output format, accuracy, operating distance, number of operators, and time required. This information is essential for understanding the technical specifications of the 3D model and its provenance. Metadata also includes a section on spatial information concerning the location of the object and its orientation in space, the dimensions of the object and the input and output formats used in the digitization process, a set of data very important for understanding the technical specifications of the digital resource and its compatibility with different software and hardware platforms. It also records the object's type, and detailed information about the acquired object's dimensions, including perimeter, area, volume, height, length, width, and altitude. Temporal coverage is also provided to specify the time period covered by the digital resource, fundamental for understanding its scope and its relevance to specific research questions.

The STARC model also allows the encoding of administrative metadata, especially of those concerning the intellectual property rights and project credits associated with the digital resources, vital for ensuring proper attribution and use of the digital resource and for complying with legal and ethical standards.

**9.4.2 The Lampadistis Monastery and the 4CH Knowledge Base**

The STARC metadata can be propagated and reused in different contexts, for example they were mapped and converted to the 4CH Ontology format and used to populate the 4CH Knowledge Base [50] to share descriptive, administrative, technical, and preservation metadata with the 4CH community of preservation and restoration experts. This process allowed the ontological description of the Lampadistis Monastery as a cultural asset and its use as an aggregating element for all the documentary resources and digital objects that refer to it, including the 3D model and the 360° Panoramic view representing it, thus forming the basis for the construction of a Digital Twin of this monumental site [51]. The transformation process from the STARC scheme to the 4CH Ontology demonstrates the quality and reusability of metadata and their ability to adapt to multiple scenarios, even the most recent semantic-based ones.

In the same framework of the 4CH initiative, additional metadata were created during the data curation and HBIM integration phase. Technical metadata were used to provide detailed information about the site's structural components, materials, and construction techniques, to support targeted analyses and conservation planning, thus enabling stakeholders to make informed decisions about preservation strategies. Additionally, administrative metadata were used to manage the extensive data gathered during the documentation phase. This metadata about data ownership, access rights, and preservation policies ensure that the data is managed effectively and in compliance with relevant regulations.

The HBIM model of the Lampadistis Monastery also includes a virtual representation of the building's spaces, constituting an additional and valuable dataset provided through the web interface. The segmentation of the model into building spaces not only enhances the user's navigation experience but also provides a more comprehensive dataset about the monument's historical evolution and architectural analysis. Each building space is represented as a virtual element that contains various metadata fields, including the building's construction date, building typology, architectural morphological elements, intervention dates, and intervention descriptions.

Furthermore, the intervention dates and intervention descriptions metadata fields provide valuable information about the monument's conservation history. These metadata fields document the various conservation interventions that have been carried out on the building over time, including the date and type of the intervention, and a description of the work carried out, essential to understand the current condition of the monument and planning future conservation actions.

The visualisation interface of HBIM implemented using the ICE Viewer enables users to view the metadata, contributes to ongoing conservation efforts, and shares insights by accessing comprehensive scientific documentation such as X-ray Fluorescence (MA-XRF) data, or viewing historical images related to the selected elements. The availability of the download of all associated data aligns with the FAIR principles and the concept of open data, while the use of preservation metadata ensures that the data remains accessible and usable over time, despite changes in technology or data formats.

**9.5 Challenges in Long-Term Preservation**

Notwithstanding the successful implementation of advanced methodologies and technologies, several challenges remain in the long-term preservation of the data of the Agios Ioannis Lampadistis Monastery, challenges that are similarly relevant for other monuments and cultural heritage sites.

**Classification**: Despite significant efforts, a fully effective solution has not yet been achieved for the classification and management of information in cultural heritage preservation projects.

This is primarily because many of the current standards are oriented towards new constructions rather than existing historical buildings.

Historical buildings require a more detailed and specific approach to documentation and classification, taking into account their unique historical and architectural characteristics. Current standards do not always provide the necessary categories or flexibility to document these unique features.

There is no global consensus on a single classification system that can be universally applied to cultural heritage. Each country tends to use its national system, which further complicates international collaboration and standardisation.

**Data Integration**: The difficulty of integrating non-standardized historical data into modern BIM and IFC systems is another significant obstacle. Historical data are often not available in digital formats compatible with these standards, making the transition complex and time-consuming.

Despite the availability of over 200 software tools capable of importing or exporting IFC files, the practical implementation of IFC for heritage data remains limited due to varying levels of software compatibility and the specific needs of heritage documentation.

The IFC schema is implemented in BIM environments through Model View Definitions (MVDs), which define subsets of the IFC schema necessary for specific data exchange requirements. The creation and implementation of MVDs are complex and require deep knowledge of the IFC schema. Furthermore, existing software must support these MVDs, which is not always the case, leading to potential interoperability issues.

**Resource Intensity**: The deployment of sophisticated digital documentation and HBIM technologies requires substantial financial, technical, and human resources. This can be a limiting factor, particularly for smaller sites or institutions with limited budgets.

**Data Management**: The vast amount of data generated requires efficient management and storage solutions. Ensuring the interoperability and long-term accessibility of these datasets poses a significant challenge, especially as technological standards evolve.

**Stakeholder Collaboration**: While collaborative platforms enhance stakeholder engagement, coordinating efforts and maintaining consistent communication among diverse groups can be complex and time-consuming.

**Multiple Platform**: Using a variety of platforms like HBIM, 3DHOP, 360° tours, and Potree for 3D modelling and long-term preservation offers significant advantages in terms of comprehensive documentation, enhanced engagement, and data security. However, it also presents challenges related to complexity, resource management, data integration, and sustainability.

Ensuring that data from different platforms are interoperable can be complex. Standards like IFC help, but there can still be challenges in integrating data seamlessly, especially when dealing with large datasets and various file formats. Inconsistent data formats and metadata standards across platforms can lead to issues in data integration and long-term usability.

Balancing these benefits and challenges requires careful planning, standardisation, and continuous investment to ensure the effective preservation of cultural heritage sites like the Agios Ioannis Lampadistis Monastery. Implementing a complete metadata-enabled long-term preservation framework is crucial. This includes efficient data management practices, flexible storage facilities, and comprehensive documentation of the site's condition, historical context, and cultural significance. Such measures support targeted analyses, effective conservation planning, and informed decision-making, ensuring the long-term sustainability of cultural heritage site.

**References**

[1] Parent, I. *et al.* (2021) 'The UNESCO/PERSIST Guidelines for the Selection of Digital Heritage for Long-Term Preservation - 2nd Edition'. Available at: https://repository.ifla.org/handle/123456789/1863 (Accessed: 26 June 2024).

[2] 'Guides to Good Practice – Archaeology Data Service' (no date). Available at: https://archaeologydataservice.ac.uk/help-guidance/guides-to-good-practice/ (Accessed: 26 June 2024).

[3] 'Digital continuity' (2022) *Wikipedia*. Available at: https://en.wikipedia.org/w/index.php?title=Digital_continuity&oldid=1126115653 (Accessed: 9 January 2024).

[4] Archives, T.N. 'The National Archives - Homepage', *The National Archives*. The National Archives. Available at: https://www.nationalarchives.gov.uk/information-management/manage-information/policy-process/digital-continuity/ (Accessed: 9 January 2024).

[5] Verburg, M., Braukmann, R. and Mahabier, W. (2023) *Making Qualitative Data Reusable - A Short Guidebook For Researchers And Data Stewards Working With Qualitative Data*. Zenodo. Available at: https://doi.org/10.5281/zenodo.8160880.

[6] Wilkinson, M.D. *et al.* (2016) 'The FAIR Guiding Principles for scientific data management and stewardship', *Scientific Data*, 3(1), p. 160018. Available at: https://doi.org/10.1038/sdata.2016.18.

[7] Flohr, P. *et al.* (2023) *Report on data management recommendations and guidelines*. Project Report 5.1. Available at: https://www.4ch-project.eu/wp-content/uploads/2024/04/D5.1-Report-on-data-management-recommendations-and-guidelines.pdf.

[8] Brinkman, L. *et al.* (2023) 'Open Science: A Practical Guide for Early-Career Researchers'. Available at: https://doi.org/10.5281/zenodo.7716153.

[9] Hardesty, J.L. *et al.* (2020) '3D Data Repository Features, Best Practices, and Implications for Preservation Models: Findings from a National Forum | Hardesty | College & Research Libraries'. Available at: https://doi.org/10.5860/crl.81.5.789.

[10] *Preservation metadata - a framework for 3D data based on the Semantic Web | IEEE Conference Publication | IEEE Xplore* (no date). Available at: https://ieeexplore.ieee.org/document/4746811 (Accessed: 26 June 2024).

[11] *PREMIS: Preservation Metadata Maintenance Activity (Library of Congress)*. Available at: https://www.loc.gov/standards/premis/ (Accessed: 26 June 2024).

[12] OAIS Reference Model (ISO 14721) (no date) OAIS Reference Model (ISO 14721). Available at: http://www.oais.info/ (Accessed: 26 June 2024).

[13] Hart, T. (2015) *Metadata Standard for Future Digital* Preservation. Available at: https://doi.org/10.13140/RG.2.2.12486.37447.

[14] *The Dublin Core Metadata Initiative (DCMI)* (2023). Available at: https://www.dublincore.org/ (Accessed: 26 June 2024).

[15] *Metadata Encoding and Transmission Standard (METS) Official Web Site | Library of Congress* (no date). Available at: https://www.loc.gov/standards/mets/ (Accessed: 26 June 2024).

[16] *PROV Model Primer* (2013) W3C. Available at: https://www.w3.org/TR/prov-primer/ (Accessed: 26 June 2024).

[17] *ARK Alliance | The Archival Resource Key (ARK)*. Available at: https://arks.org/ (Accessed: 26 June 2024).

[18] *The London Charter* (no date) *The London Charter*. Available at: https://www.london-charter.org/ (Accessed: 26 June 2024).

[19] Denard, H. (2012) 'A New Introduction to The London Charter', in *Paradata and Transparency in Virtual Heritage*. Routledge.

[20] D'Andrea, A. and Fernie, K. (2013) 'CARARE 2.0: A metadata schema for 3D cultural objects', in *2013 Digital Heritage International Congress (DigitalHeritage)*. *2013 Digital Heritage International Congress (DigitalHeritage)*, pp. 137–143. Available at: https://doi.org/10.1109/DigitalHeritage.2013.6744745.

[21] Blundell, J. *et al.* (2020) 'Metadata Requirements for 3D Data'. Available at: https://scholarshare.temple.edu/handle/20.500.12613/6751 (Accessed: 26 June 2024).

[22] Huvila, I. (2022) 'Improving the usefulness of research data with better paradata', *Open Information Science*, 6(1), pp. 28–48. Available at: https://doi.org/10.1515/opis-2022-0129..

[23] Medici, M. and Fernie, K. (2022) *Report on standards, procedures and protocols*. Project Report 4.1. Zenodo. Available at: https://zenodo.org/records/7701529 (Accessed: 26 June 2024).

[24] Moore, J., Rountrey, A. and Scates Kettler, H. (eds) (2022) *3d data creation to curation: community standards for 3d data preservation*. Chicago: Association of College and Research Libraries.

[26] Immonen, V. (2022) '3D Modelling of Heritage Objects: Representation, Engagement and Performativity of the Virtual Realm', in A. Schwan and T. Thomson (eds) *The Palgrave Handbook of Digital and Public Humanities*. Cham: Springer International Publishing, pp. 377–396. Available at: https://doi.org/10.1007/978-3-031-11886-9_20.

[27] *Basic principles and tips for 3D digitisation of cultural heritage | Shaping Europe's digital future* (2020). Available at: https://digital-strategy.ec.europa.eu/en/library/basic-principles-and-tips-3d-digitisation-cultural-heritage (Accessed: 26 June 2024).

[28] Lochhead, I. and Hedley, N. (2021) 'Designing Virtual Spaces for Immersive Visual Analytics', *KN - Journal of Cartography and Geographic Information*, 71(4), pp. 223–240. Available at: https://doi.org/10.1007/s42489-021-00087-y.

[29] Biljecki, F., Ledoux, H. and Stoter, J. (2016) 'An improved LOD specification for 3D building models', *Computers, Environment and Urban Systems*, 59, pp. 25–37. Available at: https://doi.org/10.1016/j.compenvurbsys.2016.04.005.

[30] López, F.J. *et al.* (2018) 'A Review of Heritage Building Information Modeling (H-BIM)', *Multimodal Technologies and Interaction*, 2(2), p. 21. Available at: https://doi.org/10.3390/mti2020021.

[31] *Model View Definitions (MVD)* (no date) *buildingSMART Technical*. Available at: https://technical.buildingsmart.org/standards/ifc/mvd/ (Accessed: 27 June 2024).

[32] Oostwegel, L.J.N. *et al.* (2022) 'Digitalization of culturally significant buildings: ensuring high-quality data exchanges in the heritage domain using OpenBIM', *Heritage Science*, 10(1), p. 10. Available at: https://doi.org/10.1186/s40494-021-00640-y.

[33] 'File formats' *DANS*. Available at: https://dans.knaw.nl/en/file-formats/ (Accessed: 26 June 2024).

[35] 'Recommended formats' *UK Data Service*. Available at: https://ukdataservice.ac.uk/learning-hub/research-data-management/format-your-data/recommended-formats/ (Accessed: 26 June 2024)..

[36] 'File formats' *ADS - Archaeology Data Service*. Available at: https://archaeologydataservice.ac.uk/help-guidance/guides-to-good-practice/data-analysis-and-visualisation/3d-models/creating-3d-data/file-formats/ (Accessed: 26 June 2024).

[37] *Study on quality in 3D digitisation of tangible cultural heritage | Shaping Europe's digital future* (2022). Available at: https://digital-strategy.ec.europa.eu/en/library/study-quality-3d-digitisation-tangible-cultural-heritage (Accessed: 26 June 2024)..

[38] *The European Committee for Standardization CEN-CENELEC*. Available at: https://www.cencenelec.eu/about-cen/ (Accessed: 26 June 2024).

[39] *Web3D Consortium | Open Standards for Real-Time 3D Communication*. Available at: https://www.web3d.org/ (Accessed: 26 June 2024).

[40] 'File formats' *DANS*. Available at: https://dans.knaw.nl/en/file-formats/ (Accessed: 26 June 2024).

[41] 'File formats' *ADS - Archaeology Data Service*. Available at: https://archaeologydataservice.ac.uk/help-guidance/guides-to-good-practice/data-analysis-and-visualisation/3d-models/creating-3d-data/file-formats/ (Accessed: 26 June 2024).

[43] Vassallo, V. et al. (2023) *Report on pilots*. 4.4.

[44] INCEPTION srl, *THIN-ICE Platform*. Available at: http://www.inceptionspinoff.com (Accessed: 25 June 2024

[45] *Saint John Lampadistis Church Complex THIN-ICE Platform*. Available at: https://thinice.arch.unife.it/Platform/ModelDetails?name=Saint-John-Lampadistis-Church-Complex&scope=general (Accessed: 24 June 2024).

[46] potree (2024) 'potree/potree'. Available at: https://github.com/potree/potree (Accessed: 27 June 2024).

[47] CNR-ISTI, *3DHOP 3D Heritage Online Presenter*. Available at: https://3dhop.net/ (Accessed: 26 June 2024)

[48] *Monastery of Agios Ioannis (St John) Lampadistis, Kalopanagiotis*, *APAC Laboratories*. Available at: https://apaclabs.cyi.ac.cy/virtual-visits/monastery-of-agios-ioannis-lampadistis (Accessed: 27 June 2024).

[49] Hermon, S., Niccolucci, F. and Ronzino, P. (2012) 'A Metadata Schema for Cultural Heritage Documentation', *Electronic Imaging & the Visual Arts : EVA 2012 Florence, 9-11 May 2012*, pp. 36–41. Available at: https://doi.org/10.1400/187333.

[50] *Final report on services and tools* (2024). 4CH 3.3. Zenodo. Available at: https://zenodo.org/records/11204151 (Accessed: 26 June 2024). The Knowledge Base is accessible at: https://www.4ch-cloud.eu/.

[51] *Monastery of Agios Ioannis Lampadistis 4CH Knowledge Base (beta)*. Available at: https://ch.cloud.cnaf.infn.it/omeka-s/s/4ch-kb/item/17 (Accessed: 26 June 2024).

# APPENDIX

| Name | Description | Domain | Country | Viewer |
|---|---|---|---|---|
| [ADS](#) | The Archaeology Data Service (ADS) is a digital repository established in 1996 at the University of York, UK. It holds a variety of archaeological data and publications, including text documents, images, 3D models, and geospatial data from excavations and surveys in the UK and beyond. Accredited by the CoreTrustSeal, ADS provides free and open access to its data for researchers, students, and the public. It also offers tools and resources to support data use and reuse, alongside guidelines for best practices in data management and publication. | archaeology | UK | N/A |
| [ARK](#) | ARK (The Archaeological Recording Kit) is an open-source, web-based toolkit for the collection, storage, and dissemination of archaeological data. It offers adaptable tools for data editing, creation, viewing, and sharing, all through a web interface. Compatible with any recording system, ARK provides a flexible framework and pre-fabricated tools tailored to project-specific needs, using industry-standard technologies (Apache/MySQL/PHP) | | | |
| [ARCHE](#) | ARCHE (A Resource Centre for the HumanitiEs) is a service aimed at offering stable and persistent hosting as well as dissemination of digital research data and resources for the Austrian humanities community. | general | AU | [3DHOP](#) |
| [DANS](#) | The DANS repository is an exclusively Dutch repository for the humanities, archaeology, geospatial sciences, and behavioral and social sciences. It is interoperable with the Netherlands Coalition for Digital Preservation (NCDD) | cultural heritage | NL | N/A |
| [ATON framework](#) | ATON is an open-source framework based on Node.js and Three.js, developed by B. Fanini (VHLab, CNR ISPC), for creating Web3D/WebXR applications. It adapts automatically to various devices and offers an API for manipulating scene-graphs and customizing event handling. Features include advanced 3D object rendering, spatial UIs, real-time collaboration, and integration with multimedia content, all without requiring installation for users. | archaeology | IT | [ATON](#) |

| Name | Description | Domain | Country | Viewer |
|---|---|---|---|---|
| [CFIR.science](#) | CFIR.science is a web-based digital research infrastructure designed to make archaeological artifacts freely accessible in the form of 3D data, images and metadata | archaeology | AU | [3DHOP](#) |
| [Corallum Fabrica](#) | Corallum Fabrica is an open-science project dedicated to the 3D structure of coral skeletons involving designers and scientists. Coral samples have been imaged using x-ray tomography and high resolution 3D models of their structures have been elaborated. | sealife | FR | [3DHOP](#) |
| [CyArk](#) | CyArk is a non-profit organization that specializes in the digital preservation of cultural heritage sites, architecture, and archaeological sites. Founded in 2003, CyArk uses advanced technologies such as laser scanning, photogrammetry, and 3D modeling to create digital models of cultural heritage sites around the world. 3D model are published on Sketchfab. | cultural heritage | US | [Sketchfab](#) |
| [Data Service for complex data in the arts and humanities](#) | 3D Data Service for complex data in the arts and humanities is a collaborative project funded by the UK's AHRC lead by the University of Brighton which is planning the development of a new national data service for UK research data. The project team includes Dr Doug Boyer of Duke University and the project hopes to build on the experience of developing MorphoSource. | cultural heritage | UK | N/A |
| [Digital Repository of Ireland](#) | The Digital Repository of Ireland (DRI) is a national digital repository for cultural and social data, collaboratively managed by institutions like the Royal Irish Academy, Trinity College Dublin, and the National University of Ireland, Galway. It houses a wide range of digital collections, including photographs, audio recordings, videos, text documents, and 3D data. DRI provides best practices for digital preservation and serves as a research hub in digital humanities, offering workshops and training programs. It is accredited by the CoreTrustSeal for trusted preservation and access | general | EI | N/A |

| Name | Description | Domain | Country | Viewer |
|---|---|---|---|---|
| [Digitizing Early Farming Cultures (DEFC)](#) | The objective of Digitizing Early Farming Cultures (DEFC) is the standardization and integration of archaeological research data from the Neolithic and Copper Age (7000 – 3000 BC) in Greece and Western Anatolia. | archaeology | AU | [3DHOP](#) |
| [Dynamic Collections](#) | The Dynamic Collections project at Lund University aims to create a 3D web infrastructure that will enhance higher education and research in archaeology. The project focuses on developing a new platform that will enable scholars to interact with 3D data in real-time, allowing for more dynamic and immersive learning experiences. The platform will feature a range of tools and resources to support research, teaching, and public engagement, including 3D models, interactive maps, and multimedia content. The project is a collaborative effort between archaeologists, computer scientists, and web developers. | archaeology | SW | [3DHOP](#) |
| [e-Navs.eu](#) | The project aims to digitise traditional shipbuilding to create a Digital Repository of Greek Historical / Traditional Boats. The 3D material will be used to build 3D shipbuilding models as well as virtual interactive experiences. | maritime heritage | GR | N/A |
| [Edition Topoi](#) | The Edition Topoi research platform serves the publication of citable research data such as 3D models, high-resolution pictures, RTI and other type of data. The content and its metadata are subject to peer review and made available on an Open Access basis. The published or publishable combination of citable research content and its technical and contextually relevant meta data is defined as Citable. The public data are generated via a cloud and can be directly connected with the individual computing environment. | cultural heritage | GM | [3DHOP](#) |

| Name | Description | Domain | Country | Viewer |
|---|---|---|---|---|
| [EpHEMERA](#) | EpHEMERA is an online platform that enables users to access and visualize 3D architectural and archaeological models through a standard web browser using the Potree viewer. It focuses on preserving endangered heritage in the southeastern Mediterranean area, categorizing models by risk type, such as natural disasters, war, conflict, or neglect. This helps prioritize vulnerable sites for preservation. EpHEMERA also offers related metadata, contextual information, and scholarly resources, enhancing its capabilities beyond visualization | architecture and archaeology | CY | [Potree](#) |
| [GB3D fossils](#) | The database covers macrofossil species held in British collections, and where the species are found in the UK. We are now also happy to consider the inclusion of the types of any other macrofossils. The 3D digital models can be downloaded in various formats. The online viewer does not work. | archaeology | UK | N/A |
| [Global Digital Heritage](#) | Global Digital Heritage (GDH) is a not-for-profit, private research and education organization dedicated to documenting, monitoring, and preserving our global cultural and natural heritage. | cultural heritage | | [3DHOP](#) |
| [heidICON](#) | heidICON is provided by Heidelberg University Library and is the "Virtual Slide Collection" in progress of organization of Heidelberg University. | general | GM | [3DHOP](#) |

| Name | Description | Domain | Country | Viewer |
|---|---|---|---|---|
| [Kompakk](#) | The multimedia online repository kompakkt, presented by Zoe Schubert (SBB, Berlin), facilitates the publication and annotation of 3D objects. Each object can be registered with a DOI and described using kompakkt's metadata format, which aligns with standards from Europeana, the Open Annotation Collaboration, and the W3C Web Annotation Data Model, and is compliant with CIDOC CRM. TIB Hannover is working on integrating Wikibase with kompakkt, and an extension of the DFG Viewer will support 3D objects stored in decentralized repositories. | general | GM | [Kompakk Viewer](#) |
| [MayaArch3D](#) | The project built a 3D virtual environment for analysing archaeological data. The platform provides different levels of user access. To address data storage and reuse, data were placed in a data repository iDAI for archaeological data, including 3D. DOIs were generated for these objects, and the metadata was mapped to CIDOC-CRM for better interoperability. At time of writing the GIS platform does not work. | archaeology | GM | [GIScene](#) |
| [Morphosource](#) | established by Duke University and now funded by the US National Science Foundation, is an example of a repository platform for 3D objects which includes a 3D viewer (the IIIF compliant Universal Viewer with an extension for 3D). It hosts more than 63 thousand 3D models of natural history, cultural heritage and scientific objects from museums, researchers and scholars. | archaeology | US | [Universal Viewer](#) |
| [National 3D Data Repository](#) | The National 3D Data Repository is the preferred backup solution for 3D data produced in the context of projects in Higher Education and Research in Digital Humanities. It benefits from a secure backup environment for 3D data provided by Huma-Num. Its development and maintenance are ensured by Archeovision. Its specifications are the result of work carried out within the Consortium 3D SHS. | archaeology | FR | N/A |

| Name | Description | Domain | Country | Viewer |
|---|---|---|---|---|
| OpenHeritage | The service is a non-profit organisation supported by CyArk, Historic Environment Scotland and the University of South Florida Libraries, all members of the Open Heritage Alliance. Together have significant repositories of legacy and on-going 3D research and documentation projects. The service allows free access to high-resolution 3D data (from laser scanning and photogrammetry) of cultural heritage sites. The service allows downloading 3D data after users registration. | general | US | Potree |
| PURE 3D | PURE3D is a three-year project funded by PDI-SSH, aimed at developing an online infrastructure for interactive Digital Heritage and Digital Humanities 3D content. The platform supports content creators (researchers, educators, cultural heritage managers) and end-users (students, academics, the public). It integrates various materials like annotations, images, videos, and data, creating a multimodal resource beyond traditional print. PURE3D addresses challenges such as file size, format, technical skills, and funding, and serves as a preservation repository for 3D projects. It also provides a framework to evaluate and advance 3D digital scholarship | cultural heritage | GM | Voyager |
| ReInHerit-Hub collection | The ReInHerit project proposes a sustainable heritage management model, creating a dynamic network of cultural heritage professionals. The Digital Hub and the collection is the central location for this network. Skecthfab is the viewer | general | CY | Sketchfab |
| Sketchfab |  | general | US | Sketchfab |
| The Arc/k | The Arc/k Project partners with citizen volunteers and non-profit organizations around the world to preserve endangered cultural heritage for present and future generations via digital formats including 3D and Virtual Reality. | cultural heritage | US | Sketchfab |

| Name | Description | Domain | Country | Viewer |
|---|---|---|---|---|
| [The Digital Archaeological Record (tDAR)](#) | The Digital Archaeological Record (tDAR) is a digital repository for archaeological data. It was established in 2007 as a joint effort between Arizona State University and Digital Antiquity, a non-profit organization dedicated to preserving archaeological data. tDAR is designed to support the discovery, use, and preservation of archaeological data, and it contains a wide range of digital resources, including site reports, images, and 3D models. The repository also provides guidance on best practices for managing and preserving digital resources in archaeology. | archaeology | US | N/A |
| [The Smithsonian Open Access](#) | The Smithsonian Open Access content includes high-resolution 2D and 3D images of collection items, as well as research datasets and collections metadata, which users can download and access in bulk. All of the Smithsonian's 19 museums, nine research centres, libraries, archives and the National Zoo contributed images or data. | general | US | [Voyager](#) |
| [University of Michigan Online Repository of Fossils (UMORF)](#) | The University of Michigan Online Repository of Fossils (UMORF) is a digital archive of fossil specimens housed at the University of Michigan. It is a collection of high-quality images and detailed information about fossils. The repository contains images and 3D models of fossils and includes specimens of plants, invertebrates, and vertebrates.<br>The repository provides tools for data analysis, visualization, and sharing. The archive also provides guidance on best practices for managing and preserving digital resources. | archaeology | US | [UMORF Viewer](#) |